\documentclass[11pt,a4paper,reqno]{amsart}%
\usepackage[latin1]{inputenc}
\usepackage{mathrsfs}
\usepackage{dsfont}
\usepackage{hyperref}
\usepackage{amsmath}
\usepackage{amssymb}
\usepackage{amsthm}
\usepackage{amsfonts}
\usepackage{amstext}
\usepackage{amsopn}
\usepackage{amsxtra}
\usepackage{mathrsfs}
\usepackage{dsfont}
\usepackage{esint}
\usepackage{pst-all}
\usepackage{pstricks}
\usepackage{graphicx}
\theoremstyle{plain}
\newtheorem{theorem}{Theorem}
\newtheorem{lemma}[theorem]{Lemma}

\theoremstyle{definition}

\theoremstyle{remark}
\newtheorem{remark}{Remark}

\numberwithin{equation}{section}
\newcommand{\dps}{\displaystyle}
\newcommand{\ii}{\infty}
\newcommand\R{{\ensuremath {\mathbb R} }}
\newcommand\bS{{\ensuremath {\mathbb S} }}
\newcommand\C{{\ensuremath {\mathbb C} }}

\newcommand\Z{{\ensuremath {\mathbb Z} }}

\renewcommand\phi{\varphi}

\newcommand{\wto}{\rightharpoonup}
\newcommand{\cS}{\mathcal{S}}

\newcommand{\cJ}{\mathcal{J}}

\newcommand{\cM}{\mathcal{M}}

\newcommand{\cK}{\mathcal{K}}
\newcommand{\cE}{\mathcal{E}}

\newcommand{\eps}{\epsilon}

\renewcommand{\epsilon}{\varepsilon}
\newcommand\pscal[1]{{\ensuremath{\left\langle #1 \right\rangle}}}
\newcommand{\norm}[1]{ \left| \! \left| #1 \right| \! \right| }
\newcommand{\tr}{{\rm Tr}\,}

\renewcommand{\geq}{\geqslant}
\renewcommand{\leq}{\leqslant}

\renewcommand{\tilde}{\widetilde}

\newcommand{\nn}{\nonumber}

\title[Existence of Hartree-Fock excited states]{Existence of Hartree-Fock excited states for atoms and molecules}

\author[M. Lewin]{Mathieu Lewin}
\address{CNRS \& CEREMADE, Universit\'e Paris-Dauphine, PSL Research University, F-75016 Paris, France} 
\email{mathieu.lewin@math.cnrs.fr}

\date{\today. \copyright~2017 by the author. This paper may be reproduced, in its entirety, for non-commercial purposes}

\begin{document}

\maketitle

\begin{abstract}
For neutral and positively charged atoms and molecules, we prove the existence of infinitely many Hartree-Fock critical points below the first energy threshold (that is, the lowest energy of the same system with one electron removed). This is the equivalent, in Hartree-Fock theory, of the famous Zhislin-Sigalov theorem which states the existence of infinitely many eigenvalues below the bottom of the essential spectrum of the $N$-particle linear Schrödinger operator. Our result improves a theorem of Lions in 1987 who already constructed infinitely many Hartree-Fock critical points, but with much higher energy. Our main contribution is the proof that the Hartree-Fock functional satisfies the Palais-Smale property below the first energy threshold. We then use minimax methods in the $N$-particle space, instead of working in the one-particle space.
\end{abstract}

\bigskip

The Hartree-Fock model is an important nonlinear approximation of the $N$-particle Schrödinger linear equation which describes the $N$ electrons in an atom or a molecule. Its successes and limitations for approximating the first eigenfunction of the $N$-body Hamiltonian (``ground state'') are well known, but its ability to describe excited states has probably been underestimated until recently. In a recent numerical work~\cite{BarGilGill-14}, a Hartree-Fock calculation has given very good predictions for 10 of the 11 first excited states of the H$_2$ molecule. For other recent works on Hartree-Fock and Kohn-Sham excited states in Chemistry, we refer for instance to~\cite{TasTheTha-13,LiLuYan-15} and the references therein. 

These recent developments suggest that Hartree-Fock excited states need further mathematical and numerical investigation. The purpose of this paper is to give a rigorous definition of these excited states, and to show that neutral and positively charged atoms and molecules have infinitely many stable Hartree-Fock excited states below the first energy threshold (that is, the lowest energy of the same system with one electron removed). 

The spectral properties of the Schrödinger linear operator describing the $N$ electrons are well known since the sixties. Hunziker, Van Winter and Zhislin have proved in~\cite{Zhislin-60,VanWinter-64,Hun-66} that the bottom of the essential spectrum is given by the lowest energy of the same molecule with one electron removed. Using this characterization of the first energy threshold, Zhislin and Sigalov have shown in~\cite{Zhislin-60,ZhiSig-65} that the $N$-particle Hamiltonian of a neutral or positively charged atom or molecule has infinitely many eigenvalues below its essential spectrum.

In a nonlinear theory such as Hartree-Fock, there is no clear equivalent of the essential spectrum. But it has recently been discovered in~\cite{Friesecke-03,Lewin-11} that the lowest Hartree-Fock energy of $N-1$ electrons plays a similar role as the bottom of the essential spectrum in the linear case. More precisely, calling $J^V(N)$ the minimum Hartree-Fock energy, it was shown in~\cite{Friesecke-03,Lewin-11} that 
$$J^V(N)<J^V(N-1)$$
is a necessary and sufficient condition for the compactness of all the minimizing sequences for $J^V(N)$. This result is based on the fundamental fact that the Hartree-Fock model is obtained by restricting the $N$-particle Hamiltonian to  the manifold of Slater determinants. The proof indeed relies on  ``geometric methods'' in the $N$-particle space, inspired of the linear case. There is no result of this kind for other similar nonlinear models, like the reduced Hartree-Fock model (where the exchange term is dropped~\cite{Solovej-91}) or Kohn-Sham models.

The first mathematical study of the Hartree-Fock model dates back to 1977, when Lieb and Simon proved in~\cite{LieSim-77} the existence of a minimizer for neutral and positively charged atoms and molecules, working in the one-particle space. Other important results on Hartree-Fock minimizers include estimates on the validity of the Hartree-Fock model for large atoms~\cite{Bach-92,Bach-93} and the proof of the ionization conjecture~\cite{Solovej-91,Solovej-03}. Minimizers for the Kohn-Sham model were constructed in~\cite{AnaCan-09}.

In~\cite{Lions-82c,Lions-87,Lions-88}, Lions was the first to construct infinitely many Hartree-Fock excited states, but those have an energy converging to 0, which does not correspond to the picture of the Zhislin-Sigalov theorem. As we have recalled, in the linear case, the eigenvalues converge to the bottom of the essential spectrum, which is negative. In this respect, the physical interpretation of Lions' critical points is not clear. To obtain his result, Lions used a bound on the Morse index of min-maxing sequences to infer their compactness. This method has then been further developed mathematically in~\cite{FanGho-92a,FanGho-94,Ghoussoub-93}, and later used in several models of quantum mechanics, including Dirac-Fock~\cite{EstSer-95,EstSer-99,Paturel-00,EstLewSer-08} and multiconfiguration theories~\cite{Lewin-02,Lewin-04a,CanGalLew-06}. 

Here we construct different critical points by working in the $N$-particle space. We think they are better candidates for representing molecular excited states since their energies converge to $J^V(N-1)<0$ and not to 0. 
We do not use the Morse index to obtain the compactness of minimizing sequences. We actually prove that the Hartree-Fock energy satisfies the usual \emph{Palais-Smale condition} below the first energy threshold $J^V(N-1)$, which allows us to use more classical techniques of nonlinear analysis. The validity of the Palais-Smale condition below the first energy threshold $J^V(N-1)$ is another property of Hartree-Fock theory, similar to the linear case, which allows us to think of $J^V(N-1)$ as the bottom of a kind of nonlinear essential spectrum.

The paper is organized as follows. After recalling the spectral properties of the $N$-particle Schrödinger model in Section~\ref{sec:Schrodinger}, we introduce the Hartree-Fock energy and discuss its minimizers in Section~\ref{sec:HF_min}. Then we state our main results for excited states in Section~\ref{sec:HF_excited_states}. The rest of the paper is devoted to the proof of our results.

\section{Main results}
\subsection{The $N$-particle Schrödinger Hamiltonian}\label{sec:Schrodinger}
Before turning to nonlinear Hartree-Fock theory, we start by recalling well known facts about the linear $N$-particle Schrödinger equation.
For simplicity, we work in $\R^3$ and neglect the spin variable, but it will be clear to the reader that our results can easily be extended to much more general situations (dimension $d\geq1$, particles with spin, pseudo-relativistic kinetic energy, etc). Let 
\begin{equation}
V,w\in L^{3/2}(\R^3,\R)+L^\ii_\epsilon(\R^3,\R),\qquad \text{with $w$ even,}
\label{eq:ass_V_w}
\end{equation}
be two real-valued potentials.\footnote{The notation $f\in L^{3/2}+L^\ii_\eps$ means that for any $\eps>0$ we can write $f=f_{3/2}+f_\ii$ with $\|f_\ii\|_{L^\ii}\leq\eps$, see~\cite{ReeSim4}. Such potentials are relative form-compact (hence infinitesimal form-bounded) perturbations of $-\Delta$ by~\cite[Sec.~X.2]{ReeSim2} and~\cite[Sec. XIII.4]{ReeSim4}.} 
We consider $N$ fermions in the external potential $V$ and interacting by pairs through the potential $w$. The corresponding $N$-particle Schrödinger Hamiltonian is defined by
\begin{equation}
H^V(N)=\sum_{j=1}^N-\Delta_{x_j}+V(x_j)+\sum_{1\leq j<k\leq N} w(x_j-x_k)
\label{eq:H_V_N}
\end{equation}
in the Hilbert space $\bigwedge_1^NL^2(\R^3,\C)$, which is the subspace of anti-symmetric functions in $L^2(\R^{3N},\C)$. Under our assumption~\eqref{eq:ass_V_w} the potentials $V$ and $w$ are infinitesimally $(-\Delta)$--form bounded, hence the quadratic form associated with the operator $H^V(N)$, defined by
\begin{align}
\cE^V(\Psi)&:=\pscal{\Psi,H^V(N)\Psi}\nn\\
&=\int_{\R^{3N}}|\nabla\Psi|^2+\int_{\R^{3N}}\bigg(\sum_{j=1}^NV(x_j)\nn\\
&\qquad\qquad +\sum_{1\leq j<k\leq N}w(x_j-x_k)\bigg)|\Psi(x_1,...,x_N)|^2\,dx_1\cdots dx_N,
\label{eq:def_cE_Sch}
\end{align}
is continuous and closed on the (antisymmetric) Sobolev space $\bigwedge_1^NH^1(\R^3,\C)$. More precisely, we have for some constant $C>0$
\begin{equation}
 \frac12\int_{\R^{3N}}|\nabla\Psi|^2-C\int_{\R^{3N}}|\Psi|^2\leq \cE^V(\Psi)\leq 2\int_{\R^{3N}}|\nabla\Psi|^2+C\int_{\R^{3N}}|\Psi|^2.
 \label{eq:coercive}
\end{equation}
In the following, we always work with the associated Friedrichs self-adjoint realization of $H^V(N)$ and call
$$E^V(N):=\min\sigma\left(H^V(N)\right)=\inf_{\substack{\Psi\in\bigwedge_1^N H^1(\R^3)\\ \norm{\Psi}_{L^2}=1}}\cE^V(\Psi)$$
the bottom of its spectrum. It is known that the essential spectrum of $H^V(N)$ is a half line in the form
$$\sigma_{\rm ess}\left(H^V(N)\right)=[\Sigma^V(N),\ii)$$
where
$$\Sigma^V(N)=\inf_{\substack{\{\Psi_n\}\subset \bigwedge_1^N H^1(\R^3)\\ \norm{\Psi_n}_{L^2}=1\\ \Psi_n\wto0}}\liminf_{n\to\ii}\cE^V(\Psi_n)$$
is the lowest possible limiting energy of sequences $\{\Psi_n\}$ tending weakly to zero. The HVZ theorem~\cite{Zhislin-60,VanWinter-64,Hun-66,ReeSim4,Lewin-11} states that such sequences $\{\Psi_n\}$ have to loose at least one particle at infinity, leading to the formula
\begin{equation}
 \Sigma^V(N)=\min_{k=1,...,N}\left\{E^V(N-k)+E^0(k)\right\}
 \label{eq:HVZ}
\end{equation}
where one minimizes over the number $k$ of particles which have been sent to infinity. These do not feel the external field $V$ anymore, hence their lowest energy is $E^0(k)$. 

In this paper we are interested in excited states, which correspond to the eigenvalues of $H^V(N)$ below the threshold $\Sigma^V(N)$. We therefore introduce the $k$th min-max level
\begin{align}
\lambda_k^V(N)&=\inf_{\substack{W\subset \bigwedge_1^NH^1(\R^3,\C)\\ \dim(W)=k}}\max_{\substack{\Psi\in W\\ \|\Psi\|_{L^2}=1}}\cE^V(\Psi)\nn\\
&=\sup_{\substack{W\subset \bigwedge_1^NH^1(\R^3,\C)\\ {\rm codim}(W)=k-1}}\inf_{\substack{\Psi\in W\\ \|\Psi\|_{L^2}=1}}\cE^V(\Psi),
\label{eq:lambda_k}
\end{align}
which is the $k$th eigenvalue of $H^V(N)$ counted with multiplicity when it exists, or is equal to $\Sigma^V(N)$ otherwise~\cite{ReeSim4}. Note that since $H^V(N)$ is real, the eigenfunctions can be chosen real, when they exist. One can indeed restrict to real-valued functions $\Psi$ in~\eqref{eq:lambda_k} without changing anything, and this is what we do in the rest of the paper.

The situation is slightly simplified for a repulsive interaction potential $w$, that is, under the additional assumption that
$$\boxed{w\geq0.}$$
In this case we have $E^0(k)=0$ for all $k\geq1$ and, therefore,
$$\Sigma^V(N)=E^V(N-1).$$
In other words, the essential spectrum starts when one particle is sent to infinity. We are in this situation for atoms and molecules in the Born-Oppenheimer approximation, for which
$$w(x)=\frac{1}{|x|},\qquad V(x)=-\sum_{m=1}^M\frac{z_m}{|x-R_m|}$$
where $z_m>0$ and $R_m\in\R^3$ are, respectively, the charges and locations of the $M$ classical nuclei in the molecule (or the atom, when there is only $M=1$ nucleus). It has been proved by Zhislin and Sigalov~\cite{Zhislin-60,ZhiSig-65} that 
$$\lambda_k^V(N)<\Sigma^V(N)=E^V(N-1),\qquad \forall k\geq1,$$
under the condition that 
$$N-1<\sum_{m=1}^Mz_m.$$
Hence neutral and positively charged molecules have infinitely many excited states below the first energy threshold $E^V(N-1)$. Our goal in this paper is to extend this fundamental result to Hartree-Fock theory. When $N\geq 1+\sum_{m=1}^Mz_m$, it has been shown in~\cite{Yafaev-76,VugZhi-77,Sigal-82} that $\lambda_k^V(N)=E^V(N-1)$ for $k$ large enough. In fact, for $N$ large enough we have $E^V(N)=E^V(N-1)$ and there are no eigenvalues at all~\cite{Lieb-84,Nam-12,Ruskai-82,Sigal-82,Sigal-84,LieSigSimThi-88,SecSigSol-90,FefSec-90b,LenLew-13}.

\subsection{Hartree-Fock theory}\label{sec:HF_min}
In Hartree-Fock theory, the $N$-particle wavefunctions $\Psi$ are assumed to be as much uncorrelated as possible, that is, they are given by a Slater determinant
\begin{equation}
\Psi(x_1,...,x_N)=\phi_1\wedge\cdots \wedge\phi_N(x_1,...,x_N):=\frac{1}{\sqrt{N!}}\det(\phi_i(x_j)).
\label{eq:Slater}
\end{equation}
Here the $N$ real-valued functions $\phi_1,...,\phi_N\in H^1(\R^3,\R)$ form an orthonormal system
$$\int_{\R^3}\phi_j\phi_k=\delta_{jk}.$$
For a given $\Psi$ in the form~\eqref{eq:Slater} the corresponding $\phi_j$'s are not unique. They can be replaced by $\sum_{k=1}^NU_{jk}\phi_k$ with $U\in O(N)$, without changing $\Psi$ (up to a sign when $\det(U)^N=-1$). 

The energy of a Hartree-Fock state~\eqref{eq:Slater} is computed to be~\cite{LieSim-77}
\begin{align}
&\cJ^V(\phi_1,...,\phi_N)\nn\\
&\ :=\cE^V(\phi_1\wedge\cdots \wedge\phi_N)\nn\\
&\ =\sum_{j=1}^N\int_{\R^3}|\nabla\phi_j|^2+V|\phi_j|^2\nn\\
&\quad+\frac12\iint_{\R^6}w(x-y)\left(\sum_{j=1}^N|\phi_j(x)|^2\sum_{k=1}^N|\phi_k(y)|^2-\bigg|\sum_{j=1}^N\phi_j(x)\phi_j(y)\bigg|^2\right)dx\,dy\nn\\
&\ =\sum_{j=1}^N\int_{\R^3}|\nabla\phi_j|^2+V|\phi_j|^2+\sum_{1\leq j< k\leq N}\iint_{\R^6}w(x-y)\left|\phi_j\wedge\phi_k(x,y)\right|^2\,dx\,dy,
\label{eq:HF_energy}
\end{align}
where we recall that all the functions are real-valued and that 
$$\phi_j\wedge\phi_k(x,y)=\frac{\phi_j(x)\phi_k(y)-\phi_j(y)\phi_k(x)}{\sqrt2}.$$

The set of Hartree-Fock states with finite kinetic energy
$$\cM:=\left\{\Psi=\phi_1\wedge\cdots \wedge\phi_N\in H^1(\R^{3N},\R),\ \pscal{\phi_j,\phi_k}=\delta_{jk}\right\}$$
is a smooth manifold in $\bigwedge_1^NH^1(\R^3,\R)$ and the unit sphere of $L^2(\R^{3N},\R)$. 
A \emph{Hartree-Fock critical point} is by definition a critical point of $\cE^V$ on the manifold $\cM$, that is, such that the restriction of $H^V(N)\Psi\in H^{-1}(\R^{3N})$ to the tangent space to $\cM$ at $\Psi$ vanishes. The tangent space is given by
\begin{multline*}
\mathcal{T}\cM_\Psi={\rm span}\bigg\{\phi_{i_1}\wedge\cdots \wedge\phi_{i_{N-1}}\wedge\psi,\qquad 1\leq i_1<\cdots <i_{N-1}\leq N\\
\psi\in H^1(\R^3,\R)\cap{\rm span}(\phi_1,...,\phi_N)^\perp\bigg\}
\end{multline*}
and contains all the possible excitations of one of the $N$ particles to the space orthogonal to the $\phi_j$'s. Computing the Gâteaux derivative in the direction where $\phi_j$ is excited into $\psi$ gives, after varying over all $\psi$, that 
\begin{equation}
 h_\Psi\phi_j\in{\rm span}(\phi_1,...,\phi_N)
 \label{eq:SCF_equation_perp}
\end{equation}
where
$$h_\Psi f:=\left(-\Delta +V+\sum_{j=1}^N|\phi_j|^2\ast w\right)f-\sum_{j=1}^N\big(({\phi_j}f)\ast w\big)\,\phi_j$$
is the \emph{one-particle mean-field Hamiltonian}~\cite{LieSim-77}. Note that this Hamiltonian does not depend on the basis set $\phi_j$ used to represent $\Psi$ as a Slater determinant, since the two nonlinear terms are invariant under rotations. From~\eqref{eq:SCF_equation_perp} there exists a (symmetric) matrix of Lagrange multipliers $\mu_{jk}$ such that 
$$h_\Psi\phi_j=\sum_{k=1}^N\mu_{jk}\phi_k,\qquad 1\leq j\leq N.$$
Applying a rotation $U\in SO(N)$ to the $\phi_j$'s, we can diagonalize the matrix $\mu_{jk}$ and we arrive at the well known \emph{Hartree-Fock equations}
\begin{equation}
\boxed{h_\Psi\phi_j=\mu_j\,\phi_j,\qquad j=1,...,N.}
\label{eq:HF_equations}
\end{equation}
This is a system of $N$ coupled nonlinear eigenvalue equations.

Before we turn to excited states, we now discuss the existence of minimizers. We call
\begin{equation}
J^V(N):=\inf_{\Psi\in\cM}\cE^V(\Psi)=\inf_{\substack{\phi_1,...,\phi_N\in H^1(\R^3,\R)\\ \pscal{\phi_j,\phi_k}=\delta_{jk}}}\cJ^V(\phi_1,...,\phi_N)
\label{eq:I_V_N}
\end{equation}
the Hartree-Fock ground state energy. Because $\cM$ is a subset of the sphere in the $N$-particle space, the Hartree-Fock ground state energy has to be greater than the first $N$-particle Schrödinger eigenvalue:
$$\boxed{E^V(N)\leq J^V(N).}$$
The following was proved in~\cite{Friesecke-03,Lewin-11}.

\begin{theorem}[Hartree-Fock ground state~\cite{Friesecke-03,Lewin-11}]\label{thm:ground_states}
Assume that 
$$V,w\in L^{3/2}(\R^3,\R)+L^\ii_\epsilon(\R^3,\R)$$
with $w$ even. Then the following are equivalent:
\begin{enumerate}
 \item[(i)] All the minimizing sequences $\{\Psi_n\}\subset\cM$ for $J^V(N)$ are precompact in $H^1(\R^{3N})$ and converge, after extraction, to a Hartree-Fock minimizer;
 
 \vspace{0.2cm}
 
 \item[(ii)] $J^V(N)<J^V(N-k)+J^0(k)$ for all $k=1,...,N$.
\end{enumerate}
\end{theorem}

The condition (ii) is a nonlinear version of the HVZ theorem recalled before in~\eqref{eq:HVZ} for the $N$-particle Hamiltonian $H^V(N)$. Although there is no clear notion of essential spectrum on the Hartree-Fock manifold $\cM$, we find that ``scattering states'' do start at the energy $\min\{J^V(N-k)+J^0(k),\ k=1,...,N\}$ obtained when some of the particles escape to infinity. Of course, when $w\geq0$ we have $J^0(k)=0$ and (ii) reduces to the simpler condition $J^V(N)<J^V(N-1)$. For atoms and molecules, it can be proved by induction that $J^V(N)<J^V(N-1)$ as soon as $N<1+\sum_{m=1}^Mz_m$, exactly like for $H^V(N)$, see~\cite{LieSim-77,Lions-87,Friesecke-03,Lewin-11}.

That we obtain finitely many conditions (ii) for the compactness of minimizing sequences was the main finding of ~\cite{Friesecke-03}, and may be surprising at first. Nonlinear models often lead to binding conditions involving a continuous parameter, in particular after using the concentration-compactness principle~\cite{Lions-84,Lions-84b,Lions-87}. For (ii) to hold, it is essential that the Hartree-Fock model is the restriction of the $N$-particle problem to the manifold $\cM$. There is no equivalent of Theorem~\ref{thm:ground_states} in reduced Hartree-Fock theory or in Kohn-Sham models.

\begin{remark}[Eigenvalues of minimizers]
For a minimizer, the Lagrange multipliers in~\eqref{eq:HF_equations} must be the $N$ first eigenfunctions of $h_\Psi$. If $w>0$, it has been shown in~\cite{BacLieLosSol-94,BacLieSol-94} that $\mu_N<\mu_{N+1}$, that is, the last shell of the mean-field operator $h_\Psi$ is always completely filled.
\end{remark}

\subsection{Hartree-Fock excited states}\label{sec:HF_excited_states}
For atoms and molecules, it seems reasonable to expect that the Hartree-Fock energy has infinitely many critical points in the interval $(J^V(N),J^V(N-1))$, as is the case for $H^V(N)$. Lions has shown in~\cite{Lions-82c,Lions-87} the existence of infinitely many Hartree-Fock critical points $\Psi^{(k)}\in\cM$, but those satisfy 
$$\lim_{k\to\ii}\cE^V(\Psi^{(k)})=0$$
and there are therefore only finitely many such states below $J^V(N-1)$. In~\cite{Leon-88}, L\'eon has studied  excited states defined by orthogonality conditions but those are in general not critical points.

Here we use methods from critical point theory \emph{in the $N$-body space} and define nonlinear minimax values which are all less or equal to $J^V(N-1)$. We get a critical point under the assumption that they are strictly below $J^V(N-1)$, which we prove is true for atoms and molecules. This is different from Lions who worked \emph{in the one-particle space} and obtained critical points with a much higher energy.

\subsubsection{Palais-Smale condition}
Our main contribution in this article is the proof that, when $w\geq0$, $\cE^V_{|\cM}$ satisfies the Palais-Smale (PS) condition on the interval $\big[J^V(N),J^V(N-1)\big)$. In~\cite[Rmk.~(5) p.~68]{Lions-87} and~\cite[Rmk~3) p.~307]{Lions-88}, Lions mentions that $\cJ^V$ does \emph{not} satisfy the PS property on $(-\ii,0)$ because some particles can be lost. It surely does not satisfy (PS) on $\big[J^V(N-1),0\big)$ but we prove it does below $J^V(N-1)$.

\begin{theorem}[Palais-Smale condition below $J^V(N-1)$]\label{thm:Palais-Smale}
Assume that $V,w\in L^{3/2}(\R^3,\R)+L^\ii_\epsilon(\R^3,\R)$ with $w$ even and, in addition, that $w\geq0$.
If $J^V(N)<J^V(N-1)$, then the Hartree-Fock energy $\cJ^V$ satisfies the Palais-Smale condition on the interval $\big[J^V(N),J^V(N-1)\big)$. That is, if we have a sequence $\Psi_n=\phi_{1,n}\wedge\cdots\wedge\phi_{N,n}\in\cM$ such that 
\begin{itemize}
\item[$\bullet$] $\cE^V(\Psi_n)\to c\in  \big[J^V(N),J^V(N-1)\big)$,
\item[$\bullet$] $h_{\Psi_n}\phi_{j,n}-\mu_{j,n}\phi_{j,n}\to0$ in $H^{-1}(\R^3)$ for all $j=1,...,N$ and some $\mu_{j,n}\in\R$,
\end{itemize}
then the sequence $\{\Psi_n\}$ is precompact in $H^1(\R^{3N})$ and converges strongly, after extraction of a subsequence, to some $\Psi=\phi_1\wedge\cdots\wedge \phi_N\in\cM$ which is a Hartree-Fock critical point. In particular, the critical sets
$$\cK_{\leq c}:=\left\{\Psi\in\cM\ :\ \cE^V(\Psi)\leq c,\ \big(\cE^V_{|\cM}\big)'(\Psi)=0\right\}$$
are all compact in $H^1(\R^{3N})$, for $c<J^V(N-1)$.
\end{theorem}

The proof of Theorem~\ref{thm:Palais-Smale} is given in Section~\ref{sec:proof_Palais-Smale} below and it relies very much on the assumption that $w\geq0$. It seems reasonable to expect that the Palais-Smale condition is verified below the energy threshold $\min\{J^V(N-k)+J^0(k),\ k=1,...,N\}$ for any potential $w$, but our proof does not allow to deal with this case. 

\begin{remark}
The quadratic form $q_A$ of a linear bounded-below self-adjoint operator $A$ satisfies the Palais-Smale condition on the unit sphere at some level $c$, if and only if $c$ does not belong to the essential spectrum of $A$. In particular, the $N$-particle energy $\cE^V$ satisfies the (PS) property on the interval $\big[E^V(N),E^V(N-1)\big)$ when $w\geq0$: if $\{\Psi_n\}$ is such that $(H^V(N)-c)\Psi_n\to0$ strongly in $H^{-1}(\R^{3N})$ with $c<\Sigma^V(N)=E^V(N-1)$, then $\{\Psi_n\}$ is precompact in $\bigwedge_1^NH^1(\R^3)$. Theorem~\ref{thm:Palais-Smale} gives the exact same property for the nonlinear Hartree-Fock model, with of course $E^V(N-1)$ replaced by $J^V(N-1)$ and the full derivative of $\cE^V$ by its projection onto the Hartree-Fock tangent plane.\qed
\end{remark}

\subsubsection{A new definition of Hartree-Fock excited states}
With the Palais-Smale property at hand, we can now use standard techniques from critical point theory on the smooth manifold $\cM$~\cite{AmbRab-73,BerLio-83b,Rabinowitz-86,Struwe}. The main point of our approach is to work in the $N$-particle space $\bigwedge_1^NH^1(\R^3,\R)$, on the contrary to Lions who worked in the space $H^1(\R^3,\R)^N$ containing $N$-tuple of one-particle functions. 

For instance, we can introduce the nonlinear minimax level
\begin{equation}
\boxed{c_k^V(N):=\inf_{\substack{f:\;\bS^{k-1}\to\cM\\ \text{continuous and odd}}}\;\sup_{\Psi\in f(\bS^{k-1})}\cE^V(\Psi)} 
\label{eq:c_k}
\end{equation}
which consists in minimizing over all the possible odd continuous images of spheres of dimension $k-1$ in $\cM$ (we will show below that there are such sets, hence the set in the infimum is not empty). These minimax levels satisfy some simple properties which we summarize in the following

\begin{lemma}[Properties of $c^V_k(N)$]\label{lem:c_k}
Assume that $V,w\in L^{3/2}(\R^3,\R)+L^\ii_\epsilon(\R^3,\R)$ with $w$ even. Then we have
\begin{enumerate}
 \item[(i)] $c_1^V(N)=J^V(N)$;
 
 \vspace{0,2cm}
 
 \item[(ii)] $c_k^V(N)\leq c_{k+1}^V(N)$ for all $k\geq1$;

 \vspace{0,2cm}
 
 \item[(iii)] $\lambda_k^V(N)\leq c_k^V(N)\leq J^V(N-1)$ for all $k\geq1$.
\end{enumerate}

\smallskip

\noindent If in addition $w\geq0$, then 
\begin{enumerate}
 \item[(iv)] $\dps \lim_{k\to\ii}c_k^V(N)=J^V(N-1).$
\end{enumerate}
\end{lemma}

The property that $c_k^V(N)\geq \lambda_k^V(N)$ is important for practical applications. It means that the Hartree-Fock critical levels are always an upper bound to the true Schrödinger eigenvalues.
Our main result is the following theorem.

\begin{theorem}[Existence of Hartree-Fock excited states]\label{thm:b_k}
Assume that $V,w\in L^{3/2}(\R^3,\R)+L^\ii_\epsilon(\R^3,\R)$ with $w$ even and, in addition, that $w\geq0$.
If 
$$c_k^V(N)< J^V(N-1)$$ 
then $c_k^V(N)$ is a critical value of $\cE^V_{|\cM}$. In addition, there exists a Hartree-Fock critical point in the level set $\{\cE^V(\Psi)=c^V_k(N),\ \Psi\in\cM\}$ which has a Morse index $\leq k-1$. The corresponding orbitals satisfy the Hartree-Fock equations
$$h_\Psi\phi_j=\mu_j\phi_j$$
where the $\mu_j\leq0$ are all less than the $(N+k-1)$th eigenvalue of $h_\Psi$.
\end{theorem}

\begin{remark}
The information on the Morse index was used by Lions in~\cite{Lions-87} to get compactness, see also~\cite{FanGho-94,Ghoussoub-93}. Here we have the Palais-Smale property and we do not need this for compactness, but the Morse information holds anyway.\qed
\end{remark}

\begin{remark}[Other minimax levels]
It is classical~\cite{AmbRab-73,BerLio-83b,Rabinowitz-86,Struwe,Ghoussoub-93} to define another nonlinear minimax level by
\begin{equation}
\boxed{b_k^V(N):=\inf_{\substack{F\subset \cM\ \text{compact}\\ \text{and symmetric},\\ \gamma(F)\geq k}}\;\sup_{\Psi\in F}\cE^V(\Psi)} 
\label{eq:b_k}
\end{equation}
where 
$$\gamma(F)=\inf\left\{j\geq1\ :\ \exists f:A\to\bS^{j-1}\ \text{odd and continuous}\right\}$$
is the Krasnosel'skii genus of the compact set $F$. By the Borsuk-Ulam theorem, we have $\gamma(\bS^{k-1})=k$, and since 
$\gamma\big(f(\bS^{k-1})\big)\geq \gamma(\bS^{k-1})=k$, we have
$$b_k^V(N)\leq c_k^V(N)$$
for all $k\geq1$. Lemma~\ref{lem:c_k} and Theorem~\ref{thm:b_k} hold exactly the same for $b_k^V(N)$, with the exception on the upper bound on the Morse index. From~\cite[Prop.~8.1]{Rabinowitz-86} we know that if $b_{k+j}^V(N)=b_k^V(N)$, then the critical set at this level has genus at least $j+1$:
$$\gamma\left(\left\{\Psi\in\cM\ :\ \cE^V(\Psi)=b_k^V(N),\ \big(\cE^V_{|\cM}\big)'(\Psi)=0\right\}\right)\geq j+1,$$
which is reminiscent of the multiplicity of degenerate eigenvalues in the linear case. 

Following~\cite{Viterbo-88,Coffman-88,Ghoussoub-93}, we could as well introduce another minimax level based on $\Z_2$-equivariant homology classes, and obtain all the same results, this time including the Morse index bound.

Finding the relation between all these possible minimax levels is probably a difficult question. The Hartree-Fock functional may have plenty of critical points, even with a given Morse index.
\qed
\end{remark}

\subsubsection{Application to atoms and molecules}
For neutral and positively charged atoms and molecules we are able to prove that $c_k^V(N)<J^V(N-1)$ for all $k\geq1$, and hence obtain infinitely many Hartree-Fock excited states below $J^V(N-1)$. 

\begin{theorem}[Existence of infinitely many Hartree-Fock excited states for atoms and molecules]\label{thm:atoms}
Assume that $V$ and $w$ are given by
$$V(x)=-\sum_{m=1}^M\frac{z_m}{|x-R_m|}\qquad\text{and}\qquad w(x)=\frac{1}{|x|},$$
where $z_m>0$ and $R_m\in\R^3$. If $N<1+\sum_{m=1}^Mz_m$, then we have
$$\boxed{c_k^V(N)<J^V(N-1)}$$
for all $k\geq1$. In particular, the Hartree-Fock energy has infinitely many critical points in the interval $\big(J^V(N),J^V(N-1)\big)$, with energies converging to $J^V(N-1)$.
\end{theorem}

This result is the Hartree-Fock equivalent of the famous Zhislin-Sigalov theorem in the linear $N$-particle Schrödinger case.

\begin{remark}[Negatively charged molecules]
When $N\geq 1+\sum_{m=1}^Mz_m$, we expect that $b_k^V(N)=c_k^V(N)=J^V(N-1)$ for $k$ large enough, similarly as in the linear case~\cite{Yafaev-76,VugZhi-77,Sigal-82}. Lieb has proved in~\cite{Lieb-84} that $J^V(N)=J^V(N-1)$ when $N\geq 1+2\sum_{m=1}^Mz_m$ (the proof, written in the Schrödinger case works the same in Hartree-Fock theory). For $M=1$, Solovej has shown in~\cite{Solovej-96,Solovej-03} that there exists a constant $C$ such that $J^V(N)=J^V(N-1)$ when $N\geq C+z_1$.\qed
\end{remark}

The rest of the paper is devoted to the proof of our results.

\section{Proof of Theorem~\ref{thm:Palais-Smale} on the Palais-Smale property}\label{sec:proof_Palais-Smale}

\subsection{Geometric properties of Hartree-Fock states}

We will freely make use of the ``geometric techniques'' for $N$-particle states introduced in~\cite{Lewin-11}, to which we refer for details (see also~\cite{Friesecke-03}). The following lemma summarizes what is needed for our purposes.

\begin{lemma}[Geometric properties of weakly convergent Hartree-Fock sequences~\cite{Lewin-11}]\label{lem:geometric}
Assume that $V,w\in L^{3/2}(\R^3,\R)+L^\ii_\epsilon(\R^3,\R)$ with $w$ even and, in addition, that $w\geq0$.
Let $\Psi_n=\phi_{1,n}\wedge\cdots \phi_{N,n}\in\cM$ be a sequence of Hartree-Fock states, such that $\pscal{\phi_{j,n},\phi_{k,n}}=\delta_{jk}$ and $\phi_{j,n}\wto \phi_j$ weakly in $H^1(\R^3)$. Let 
$$\Psi:=\phi_1\wedge\cdots \wedge\phi_N$$
be the weak limit of $\{\Psi_n\}$ in $H^1(\R^{3N})$, which is such that $\int_{\R^{3N}}|\Psi|^2\leq1$. Then we have
\begin{equation}
\boxed{ \liminf_{n\to\ii}\cE^V(\Psi_n)\geq \big(1-\|\Psi\|^2\big)\,J^V(N-1)+\cE^V(\Psi).}
 \label{eq:weak_limit_geometric}
\end{equation}
\end{lemma}

\begin{remark}
Without the assumption on the sign of $w$, it was proved in~\cite{Lewin-11} that
\begin{equation}
 \liminf_{n\to\ii}\cE^V(\Psi_n)\geq \big(1-\|\Psi\|^2\big)\,\min_{k=1,...,N}\big\{J^V(N-k)+J^0(k)\big\}+\cE^V(\Psi)
 \label{eq:weak_limit_geometric_nosign}
\end{equation}
but the argument is longer and this more general inequality is not needed for our purposes.
Note that Theorem~\ref{thm:ground_states} follows immediately from~\eqref{eq:weak_limit_geometric_nosign} (from~\eqref{eq:weak_limit_geometric} when $w\geq0$), using that $\cE^V(\Psi)\geq \|\Psi\|^2J^V(N)$ since $\Psi/\|\Psi\|^2$ is a Slater determinant.\qed
\end{remark}

\begin{proof}
Since $V$ is relatively $(-\Delta)$--form-compact under our assumptions, we find
$$\lim_{n\to\ii}\int_{\R^3}V|\phi_{j,n}|^2=\int_{\R^3}V|\phi_{j}|^2,\qquad j=1,...,N$$
see, e.g.,~\cite[Lem.~1.2]{Friesecke-03} and~\cite[Lem.~5]{Lewin-11}. The other terms being weakly lower continuous by Fatou's lemma (since $w\geq0$ by assumption), we find from~\eqref{eq:HF_energy} that
\begin{multline}
 \liminf_{n\to\ii}\cE^V(\Psi_n)\geq \sum_{j=1}^N\int_{\R^3}|\nabla\phi_j|^2+V|\phi_j|^2\\+\sum_{1\leq j< k\leq N}\iint_{\R^6}w(x-y)\left|\phi_j\wedge\phi_k(x,y)\right|^2\,dx\,dy.
 \label{eq:wlsc}
\end{multline}
It is not easy to interpret the right side of~\eqref{eq:wlsc} since the weak limits $\phi_j$ do not necessarily form an orthonormal system. Applying an appropriate ($n$-independent) rotation to the $\phi_{j,n}$, we may always assume that $\int_{\R^3}\phi_j\phi_k=0$ for $j\neq k$. The functions are however not necessarily normalized and we only have $\|\phi_j\|_{L^2}\leq1$. For such an orthogonal system, a simple but tedious calculation gives 
\begin{multline}
\cE^V(\Psi)=\sum_{j=1}^N\prod_{\ell\neq j}\|\phi_\ell\|^2\int_{\R^3}|\nabla\phi_j|^2+V|\phi_j|^2\\+\sum_{1\leq j< k\leq N}\prod_{\ell\notin\{j,k\}}\|\phi_\ell\|^2\iint_{\R^6}w(x-y)\left|\phi_j\wedge\phi_k(x,y)\right|^2\,dx\,dy,
 \label{eq:Energy_Psi}
\end{multline}
which is different from~\eqref{eq:wlsc}. We now have to study the difference and prove that it can be bounded from below by $(1-\|\Psi\|^2)J^V(N-1)$. 

This is exactly what has been done in~\cite{Lewin-11}. In~\cite[Lemma~10]{Lewin-11} it is shown that the right side of~\eqref{eq:wlsc} can be written in the form
$$\text{r.h.s. of~\eqref{eq:wlsc}} = \sum_{n=1}^{N-1}\tr\big(H^V(n)G_n\big)+\cE^V(\Psi)$$
where each $G_n=(G_n)^*\geq0$ is a convex combination of $n$-particle Hartree-Fock states and
$$G_0+\sum_{n=1}^{N-1}\tr(G_n)+\|\Psi\|^2=1,$$
with $0\leq G_0\leq1$. More precisely, the operators $G_n$ are given by
\begin{multline}
G_n=\\\sum_{I=\{i_1<\cdots< i_n\}\subset\{1,...,N\}}\prod_{\ell\in\{1,...,N\}\setminus I}\left(1-\norm{\phi_{\ell}}^2\right)\big|\phi_{i_1}\wedge\cdots\wedge\phi_{i_n}\big\rangle\big\langle\phi_{i_1}\wedge\cdots\wedge\phi_{i_n}\big|,
\label{eq:convex_comb_HF}
\end{multline}
see~\cite[Example~16]{Lewin-11}. We have
$$\tr\big(H^V(n)G_n\big)\geq J^V(n)\tr(G_n)$$
since $G_n$ is the convex combination~\eqref{eq:convex_comb_HF} of $n$-particle Slater determinants and $J^V(n)$ is the lowest $n$-particle Hartree-Fock energy. Using that $0\geq J^V(n)\geq J^V(N-1)$ for $n\leq N-1$, we get~\eqref{eq:wlsc} as stated.
\end{proof}

\subsection{Proof of Theorem~\ref{thm:Palais-Smale}}
From the assumption that $\cE^V(\Psi_n)\to c$, we deduce that $\{\Psi_n\}$ is bounded in $H^1(\R^{3N})$ and hence that $\{\phi_{j,n}\}$ is bounded in $H^1(\R^3)$ for $j=1,...,N$. We then have
$$-\frac{\Delta}{2}-C\leq h_{\Psi_n}\leq -2\Delta+C$$
where $C$ is independent of $n$, and therefore 
$$\pscal{\phi_{j,n},h_{\Psi_n}\phi_{j,n}}=\mu_{j,n}+o(1)$$
are uniformly bounded for all $j=1,...,N$. Up to extraction of a subsequence, we may assume that $\mu_{j,n}= \mu_j$ is independent of $n$, and that $\phi_{j,n}\wto \phi_j$ weakly in $H^1(\R^3)$. We denote by
$$\Psi=\phi_1\wedge\cdots \wedge\phi_N$$
the weak limit of $\Psi_n$, as in Lemma~\ref{lem:geometric}. From~\eqref{eq:wlsc} we see that
\begin{equation}
c\geq \big(1-\|\Psi\|^2\big)\,J^V(N-1)+\cE^V(\Psi).
\label{eq:lower_bound_c}
\end{equation}
If $\Psi=0$ we already get a contradiction with $c<J^V(N-1)$, so we may assume that $\Psi\neq0$, which implies that $\phi_j\neq0$ for all $1\leq j\leq N$. Our goal is to prove that $\cE^V(\Psi)\geq c\|\Psi\|^2$ in order to conclude that $\|\Psi\|=1$. For this we have to use the fact that $\Psi_n$ is almost a critical point of $\cE^V$ on $\cM$.

Passing to the weak limit in the Hartree-Fock equation, we get
\begin{equation}
 h_\Psi\phi_j=\mu_j\phi_j,\qquad j=1,...,N.
 \label{eq:SCF_limit}
\end{equation}
This implies that the $\phi_j$ corresponding to distinct eigenvalues $\mu_j$ must be orthogonal to each other. If several $\mu_j$ coincide, then we can apply an $n$-independent rotation to the corresponding $\phi_{j,n}$ (this preserves the property that $h_{\Psi_n}\phi_{j,n}-\mu_j\phi_{j,n}\to0$), and  assume that the limiting $\phi_j$ are orthogonal to each other. So, from now on we assume that $\pscal{\phi_j,\phi_k}=0$ for $k\neq j$. Of course, we \emph{a priori} only have $0<\|\phi_j\|\leq 1$ and our goal is to show that $\|\phi_j\|=1$ for all $j=1,...,N$.

Taking the scalar product with $\phi_j$ in~\eqref{eq:SCF_limit} we obtain
$$\int_{\R^3}|\nabla\phi_j|^2+V|\phi_j|^2+\sum_{k=1}^N\iint_{\R^6}w(x-y)|\phi_j\wedge\phi_k(x,y)|^2\,dx\,dy=\mu_j\norm{\phi_j}^2.$$
Summing over $j$ and symmetrizing the double sum, we infer that
\begin{multline}
\sum_{j=1}^N\|\phi_j\|^{-2}\int_{\R^3}|\nabla\phi_j|^2+V|\phi_j|^2\\
=\sum_{j=1}^N\mu_j-\sum_{1\leq j<k\leq N}\left(\|\phi_j\|^{-2}+\|\phi_k\|^{-2}\right)\iint_{\R^6}w(x-y)|\phi_j\wedge\phi_k(x,y)|^2\,dx\,dy. 
\label{eq:relation_equation}
\end{multline}
Since 
$\norm{\Psi}^2=\prod_{j=1}^N\|\phi_j\|^2$,
we have, inserting~\eqref{eq:relation_equation} in~\eqref{eq:Energy_Psi} ,
\begin{align*}
\frac{\cE^V(\Psi)}{\|\Psi\|^2}= &\sum_{j=1}^N\|\phi_j\|^{-2}\int_{\R^3}|\nabla\phi_j|^2+V|\phi_j|^2\\
&\qquad+\sum_{1\leq j<k\leq N}\|\phi_j\|^{-2}\|\phi_k\|^{-2}\iint_{\R^6}w(x-y)|\phi_j\wedge\phi_k(x,y)|^2\,dx\,dy\\
= &\sum_{j=1}^N\mu_j-\sum_{1\leq j<k\leq N}\iint_{\R^6}w(x-y)|\phi_j\wedge\phi_k(x,y)|^2\,dx\,dy\\
&+\!\!\sum_{1\leq j<k\leq N}\left(\|\phi_j\|^{-2}-1\right)\left(\|\phi_k\|^{-2}-1\right)\iint_{\R^6}w(x-y)|\phi_j\wedge\phi_k(x,y)|^2\,dx\,dy\\
\geq &\sum_{j=1}^N\mu_j-\sum_{1\leq j<k\leq N}\iint_{\R^6}w(x-y)|\phi_j\wedge\phi_k(x,y)|^2\,dx\,dy.
\end{align*}
In the last line we have used that $\|\phi_j\|\leq 1$ and that $w\geq0$. Using now our assumption that 
$$h_{\Psi_n}\phi_{j,n}-\mu_j\phi_{j,n}\to0\qquad\text{strongly in $H^{-1}(\R^3)$},$$
we find after similar manipulations
\begin{align*}
c=&\lim_{n\to\ii}\cE^V(\Psi_n)\\
=&\lim_{n\to\ii}\bigg(\sum_{j=1}^N\int_{\R^3}|\nabla\phi_{j,n}|^2+V|\phi_{j,n}|^2\\
&\qquad +\sum_{1\leq j<k\leq N}\iint_{\R^6}w(x-y)|\phi_{j,n}\wedge\phi_{k,n}(x,y)|^2\,dx\,dy\bigg)\\
=&\sum_{j=1}^N\mu_j-\lim_{n\to\ii}\sum_{1\leq j<k\leq N}\iint_{\R^6}w(x-y)|\phi_{j,n}\wedge\phi_{k,n}(x,y)|^2\,dx\,dy\\
\leq&\sum_{j=1}^N\mu_j-\sum_{1\leq j<k\leq N}\iint_{\R^6}w(x-y)|\phi_{j}\wedge\phi_{k}(x,y)|^2\,dx\,dy.
\end{align*}
In the last line we have used Fatou's Lemma and the assumption that $w\geq0$. At this point we have shown that 
$$\frac{\cE^V(\Psi)}{\|\Psi\|^2}\geq c,$$
as desired. 

Coming back to~\eqref{eq:lower_bound_c}, we get
$$c\geq \big(1-\|\Psi\|^2\big)\,J^V(N-1)+\cE^V(\Psi)\geq \big(1-\|\Psi\|^2\big)\,J^V(N-1)+\|\Psi\|^2c.$$
Due to our assumption that $c<J^V(N-1)$, this implies $\cE^V(\Psi)=c$ and $\|\Psi\|=1$ hence that $\Psi_n\to\Psi$ strongly in $L^2(\R^{3N})$. Similarly, we have $\phi_{j,n}\to\phi_j$ strongly in $L^2(\R^3)$, hence in $L^p(\R^3)$ for all $2\leq p<6$, by the Sobolev inequality. This strong convergence implies that we can pass to the limit in the quartic term involving $w$. We then deduce from the fact that $\cE^V(\Psi_n)\to\cE^V(\Psi)$ that
$$\lim_{n\to\ii}\int_{\R^{3N}}|\nabla\Psi_n|^2=\int_{\R^{3N}}|\nabla\Psi|^2.$$
This gives the sought-after strong convergence $\Psi_n\to\Psi$ in $H^1(\R^{3N})$. Similarly, $\phi_{j,n}\to\phi_j$ strongly in $H^1(\R^3)$ and we already know that $h_\Psi\phi_j=\mu_j\phi_j$, that is, $\Psi$ is a Hartree-Fock critical point.
\qed

\section{Proof of Lemma~\ref{lem:c_k} on the properties of $c_k^V(N)$}
Since (i) and (ii) are obvious from the definition, we only show that
\begin{enumerate}
 \item[(iii)] $\lambda_k^V(N)\leq c_k^V(N)\leq J^V(N-1)$ for all $k\geq1$,
\end{enumerate}
and
\begin{enumerate}
 \item[(iv)] $\dps\lim_{k\to\ii}c_k^V(N)= J^V(N-1)$, when $w\geq0$.
\end{enumerate}
For completeness, we also prove the corresponding results for the minimax levels $b_k^V(N)$ defined in~\eqref{eq:b_k} using the Krasnosel'skii genus.

\subsubsection*{Proof that $c_k^V(N)\geq b_k^V(N)\geq \lambda_k^V(N)$.} It is well known that in the linear case all the minimax levels coincide, that is,
\begin{equation}
 \lambda_k^V(N)=\inf_{\substack{f:\;\bS^{k-1}\to\cS\\ \text{continuous and odd}}}\;\sup_{\Psi\in f(\bS^{k-1})}\cE^V(\Psi)=\inf_{\substack{F\subset \cS\text{ compact}\\ \text{and symmetric},\\ \gamma(F)\geq k}}\;\sup_{\Psi\in F}\cE^V(\Psi),
 \label{eq:lambda_minimax}
\end{equation}
where 
$$\cS=\left\{\Psi\in\bigwedge_1^NH^1(\R^3,\R)\ :\ \|\Psi\|_{L^2}=1\right\}$$
is the class of fermionic $N$-particle states with finite kinetic energy. Then obviously 
$$c_k^V(N)\geq b_k^V(N)\geq\lambda_k^V(N).$$ 
To prove~\eqref{eq:lambda_minimax}, we take the unit sphere of any $k$-dimensional space in $\bigwedge_1^NH^1(\R^3)$ and deduce from the usual min-max principle that $\lambda_k^V(N)$ is larger or equal to the first minimax on the right side (which is itself larger than the last). Conversely, if $F$ has genus $\gamma(F)\geq k$, then it must intersect any space of co-dimension $k-1$ by~\cite[Prop.~7.8]{Rabinowitz-86}. This implies the reverse inequality, by the linear max-min formula
$$\lambda_k^V(N)=\sup_{\substack{W\subset \bigwedge_1^NH^1(\R^3,\C)\\ {\rm codim}(W)=k-1}}\inf_{\substack{\Psi\in W\\ \|\Psi\|_{L^2}=1}}\cE^V(\Psi).$$

\subsubsection*{Proof that $c_k^V(N)\leq J^V(N-1)$.}
Let $\phi_1,...,\phi_{N-1}$ be a system of $N-1$ orthonormal functions in $H^1(\R^3)$ with support in the ball $B_{R_0}$ of radius $R_0$. Let $E$ be an arbitrary space of dimension $k$ in $C^\ii_c(B_1)$, with orthonormal basis $\psi_1,...,\psi_k$. Denote by $E_R:=\{R^{-3/2}\psi(x/R-2\nu),\ \psi\in V\}$ its dilate by a coefficient $R$, and translate in the direction $\nu\in\bS^2$ at the distance $2R$. When $R\geq R_0$, the functions in $E_R$ have a support disjoint from that of the $\phi_j$'s. Let $\psi_{R,j}(x)=R^{-3/2}\psi_j(x/R-2\nu)$ be the corresponding basis. We then look at the continuous odd map
\begin{align*}
f:\bS^{k-1}&\to \cM \\
 \omega&\mapsto \phi_1\wedge\cdots \wedge\phi_{N-1}\wedge\psi_R(\omega),
\end{align*}
where
$$\psi_R(\omega)=\sum_{j=1}^k\omega_j\,\psi_{R,j}$$
runs over the whole unit sphere of $E_R$.
Since $\psi_R(\omega)$ has a support disjoint from that of the $\phi_j$'s for $R\geq R_0$, we obtain after a calculation that
\begin{multline}
c_k^V(N)\leq \cE^V(\phi_1\wedge\cdots \wedge\phi_{N-1})+\max_{\substack{\psi\in E_R\\ \|\psi\|=1}}\bigg(\int_{\R^3}|\nabla\psi|^2+V\psi^2\\
+\sum_{j=1}^{N-1}\iint_{\R^6}w(x-y)\phi_j(x)^2\psi(y)^2\,dx\,dy\bigg).
\label{eq:estim_sup_c_k}
\end{multline}
Since the term in the parenthesis on the right tends to $0$ when $R\to\ii$, we conclude that 
$$c_k^V(N)\leq \cE^V(\phi_1\wedge\cdots \wedge\phi_{N-1}).$$
Optimizing over $\phi_1,...,\phi_{N-1}$ (and taking $R_0\to\ii$) finally gives the claimed inequality $c_k^V(N)\leq J^V(N-1)$.

\subsubsection*{Proof of the convergence to $J^V(N-1)$ when $w\geq0$.}
Since $b_k^V(N)\leq c_k^V(N)\leq J^V(N-1)$, it suffices to show that 
$$\liminf_{k\to\ii}b_k^V(N)\geq J^V(N-1).$$
For this, we fix an arbitrary orthonormal basis $\{\Psi_j\}_{j\geq1}$ of the $N$-particle space $\bigwedge_1^NL^2(\R^3,\R)$. Let $F_k$ be a symmetric compact set in $\cM$, such that $\gamma(F_k)\geq k$ and 
$$\max_{F_k}\;\cE^V\leq b_k^V(N)+\frac{1}{k}.$$
From the intersection property~\cite[Prop.~7.8]{Rabinowitz-86}, we know that $F_k$ must intersect $\{\Psi_1,...,\Psi_{k-1}\}^\perp$, at a point that we call $\tilde\Psi_k\in F_k$. The sequence $\{\tilde\Psi_k\}\subset\cM$ then satisfies $\tilde\Psi_k\wto0$ weakly in $\bigwedge_1^NL^2(\R^3)$ and
$$\cE^V(\tilde\Psi_k)\leq \max_{F_k}\;\cE^V\leq b_k^V(N)+\frac{1}{k}.$$
In particular, $\tilde\Psi_k$ is bounded in $H^1(\R^{3N})$ and converges weakly to 0 in that space. Since $w\geq0$ and $\tilde\Psi_k\wto0$, we infer from~\eqref{eq:weak_limit_geometric} in Lemma~\ref{lem:geometric} that 
$$\liminf_{k\to\ii}\cE^V(\tilde\Psi_k)\geq J^V(N-1),$$
hence $J^V(N-1)\leq \liminf_{k\to\ii}b_k^V(N)$, as we wanted.\qed

\section{Proof of Theorem~\ref{thm:b_k}}
The proof follows from classical arguments in critical point theory. We use for instance~\cite[Thm.~(4) p.~53]{Ghoussoub-91}, but there are many other related results of the same kind~\cite{Bahri-81,Hofer-85,BahLio-85,BahLio-88,Lions-87,LazSol-88,Viterbo-88,Solimini-89,Tanaka-89,Ghoussoub-93,FanGho-94,Fang-95}. From the Palais-Smale property shown in Theorem~\ref{thm:Palais-Smale} and the assumption that $c_k^V(N)<J^V(N-1)$, we obtain the existence of a critical point at the level $c_k^V(N)$, with a Morse index $\leq k-1$. 
Computing the second-order variation when only one $\phi_j$ is moved, we find as in~\cite{Lions-87} and~\cite{Ghoussoub-93} that
$$
\pscal{\psi,(h_\Psi-\mu_j)\psi}\geq0
$$
for all $\psi$ which is both orthogonal to $\phi_1,...,\phi_N$ and in a space of codimension $\leq k-1$. In total this must hold in a space of codimension $\leq N+k-1$. By~\cite[Lemma~II.2]{Lions-87}, this shows that $\mu_j$ is less than the $(N+k-1)$th eigenvalue of $h_\Psi$, as we claimed.\qed

\section{Proof of Theorem~\ref{thm:atoms}}

We know that there is a Hartree-Fock minimizer $\Psi=\phi_1\wedge\cdots \wedge\phi_{N-1}$ for $J^V(N-1)$. The corresponding orbitals satisfy
$$h_\Psi \phi_j=\mu_j\phi_j$$
where $\mu_1\leq\cdots\leq\mu_{N-1}<0$ are the $N-1$ first eigenvalues of the one-particle mean-field operators $h_\Psi$, see~\cite{LieSim-77,Lions-87}. The strict negativity of $\mu_{N-1}$ gives that the functions $\phi_j$ decay exponentially fast at infinity, as well as their gradient~\cite[Theorem~3.2]{LieSim-77}. So we can truncate the $\phi_j$'s and, after orthonormalization, we obtain an orthonormal family $\phi_j^{(R)}$ with support in a ball $B_{R/2}$, such that 
$$\cE^V(\phi_1^{(R)}\wedge\cdots\wedge \phi_{N-1}^{(R)})\leq J^V(N-1)+O(e^{-aR})$$
and
$$\sum_{j=1}^N\norm{\phi_j^{(R)}-\phi_j}_{H^1(\R^3)}=O(e^{-aR}).$$
We then use~\eqref{eq:estim_sup_c_k} with these functions and $R_0=R/2$. We obtain
\begin{multline}
c_k^V(N)\leq J^V(N-1)+O(e^{-aR}) +\max_{\substack{\psi_R\in E_R\\ \|\psi_R\|=1}}\bigg(\int_{\R^3}|\nabla\psi_R|^2+V\psi_R^2\\
+\sum_{j=1}^{N-1}\iint_{\R^6}\frac{\phi_j^{(R)}(x)^2\psi_R(y)^2}{|x-y|}\,dx\,dy\bigg).
\label{eq:estim_sup_c_k_atoms}
\end{multline}
For $\psi_R(x)=R^{-3/2}\psi(x/R-2\nu)$ with $\psi\in E$, we have by scaling 
$$\int_{\R^3}|\nabla\psi_R|^2=\frac1{R^2}\int_{\R^3}|\nabla\psi|^2$$
and
$$\int_{\R^3}\left(V+\sum_{j=1}^{N-1}|\phi_j^{(R)}|^2\ast\frac{1}{|\cdot|}\right)\psi_R^2=\frac{N-1-Z}{2R}\int_{\R^3}\frac{\psi(x)^2}{|x|}\,dx+O\left(\frac{1}{R^2}\right)$$
where $Z=\sum_{m=1}^Mz_m$. So we obtain
$$
c_k^V(N)\leq J^V(N-1)+\frac{N-1-Z}{2R}\min_{\substack{\psi\in E\\ \|\psi\|_{L^2}=1}}\int_{\R^3}\frac{\psi(x)^2}{|x|}\,dx+O\left(\frac1{R^2}\right).
$$
This is negative for $R$ large enough, under our assumption that $N<Z+1$, since $E$ is finite-dimensional.\qed

\subsection*{Acknowledgement} 
I thank \'Eric S\'er\'e for useful comments.
This project has received funding from the European Research Council (ERC) under the European Union's Horizon 2020 research and innovation programme (grant agreement MDFT No 725528).


\end{document}